\documentclass[12pt,a4paper]{article}
\usepackage{graphics}
\begin{document}
\baselineskip 0.25in
\title{\Huge{Parallel Evaluation of Mathematica Programs in Remote 
             Computers Available in Network}}
\author{\Large {Santanu K. Maiti} \\ \\
         \Large {E-mail: {\em santanu.maiti@saha.ac.in}} \\ \\
          \Large {Theoretical Condensed Matter Physics Division} \\
           \Large {Saha Institute of Nuclear Physics} \\
            \Large {1/AF, Bidhannagar, Kolkata-700 064, India}}
\date{}
\maketitle
\newpage
\tableofcontents

\newpage
\begin{center}
\addcontentsline{toc}{section}{\bf {Preface}}
{\Large \bf Preface}
\end{center}
Mathematica is a powerful application package for doing mathematics and
is used almost in all branches of science. It has widespread applications
ranging from quantum computation, statistical analysis, number theory,
zoology, astronomy, and many more. Mathematica gives a rich set of 
programming extensions to its end-user language, and it permits us to 
write programs in procedural, functional, or logic (rule-based) style, 
or a mixture of all three. For tasks requiring interfaces to the external 
environment, mathematica provides mathlink, which allows us to communicate 
mathematica programs with external programs written in C, C++, F77, F90, 
F95, Java, or other languages. It has also extensive capabilities for 
editing graphics, equations, text, etc.

In this article, we explore the basic mechanisms of parallelization of a 
mathematica program by sharing different parts of the program into all 
other computers available in the network. Doing the parallelization, we
can perform large computational operations within a very short period of
time, and therefore, the efficiency of the numerical works can be achieved.
Parallel computation supports any version of mathematica and it also works
as well even if different versions of mathematica are installed in different
computers. The whole operation can run under any supported operating system
like Unix, Windows, Macintosh, etc. Here we focus our study only for the
Unix based operating system, but this method works as well for all other
cases.

\newpage
\section{Introduction}

Mathematica, a system of computer programs, is a high-level computing
environment including computer algebra, graphics and programming. 
Mathematica is specially suitable for mathematics, since it incorporates
symbolic manipulation and automates many mathematical operations.
The key intellectual aspect of Mathematica is the invention of a new kind 
of symbolic computation language that can manipulate the very wide range of 
objects needed to achieve the generality required for technical computing
by using a very small number of basic primitives. Mathematica is now 
emerging as an important tool in many branches of computing, and today it 
stands as the world's best system for general computation.

Parallelization is a form of computation in which one can perform many
operations simultaneously. Parallel computation uses multiple processing 
elements simultaneously to finish a particular job. This is accomplished 
by breaking the job into independent parts so that each processing element 
can execute its part of the algorithm simultaneously with the others. The 
processing elements can be diverse and include resources such as a single 
computer with multiple processors, several networked computers, specialized 
hardware, or any combination of the above.

In this article, we narrate the basic mechanisms for parallelizing a 
mathematica program by running its independent parts in several computers
available in the network. Since all the basic mathematical operations are 
performed quite nicely in any version of mathematica, it does not matter 
even if different versions of mathematica are installed in different computers
those are required for the parallel computing. For our illustrative purposes,
here we describe the parallelization technique for the Unix based operating
system only.

\section{How to Open Slaves in Local Computer ?}

In parallel computation, different segments of a job are computed 
simultaneously. These operations can be performed either in a local 
computer or in remote computers available in the network. Separate 
operations are exhibited in separate mathematica slaves. In order to
emphasize the basic mechanisms, let us now describe the way of starting
a mathematica slave in a local computer. To do this, first we load the
following package in a mathematica notebook.

\vskip 0.2in
\begin{center}
{\fbox{\parbox{2.25in}{\centering{
Needs[``Parallel\`{}Parallel\`{}"]
}}}}
\end{center}
To enable optional features, then we load the package,

\vskip 0.2in
\begin{center}
{\fbox{\parbox{2.45in}{\centering{
Needs[``Parallel\`{}Commands\`{}"]
}}}}
\end{center}
Now we can open a mathematica slave in the local computer by using the 
command,

\vskip 0.2in
\begin{center}
{\fbox{\parbox{4.25in}{\centering{
LaunchSlave[``localhost", ``math -noinit -mathlink"]
}}}}
\end{center}
Using this command, several mathematica slaves can be started from the
master slave. Now it becomes much more significant if we specify the 
names of different slaves so that independent parts of a job can be shared
into different slaves appropriately. For our illustrations, below we
give some examples how different slaves can be started with specific 
names.

\vskip 0.2in
\begin{center}
{\fbox{\parbox{4.25in}{\centering{
link1=LaunchSlave[``localhost", ``math -noinit -mathlink"] \\
link2=LaunchSlave[``localhost", ``math -noinit -mathlink"] \\
link3=LaunchSlave[``localhost", ``math -noinit -mathlink"] 
}}}}
\end{center}
Here link1, link2 and link3 correspond to the three different slaves. 
The details of these slaves can be available by using the following 
command,

\vskip 0.2in
\begin{center}
{\fbox{\parbox{5.25in}{\centering{
TableForm[RemoteEvaluate[$\{\$$ProcessorID, $\$$MachineName, $\$$SystemID,
$\$$ProcessID, $\$$Version$\}$], TableHeadings$\rightarrow$$\{$None,$\{$``ID",
``host", ``OS", ``process", ``Mathematica Version"$\}\}$]
}}}}
\end{center}
The output of the above command becomes (as an example),
\vskip 0.2in
\begin{center}
\begin{tabular}{|c|c|c|c|c|}
\hline
ID & host & OS & process & Version \\ \hline 
1 & tcmpibm & AIX-Power64 & 463002 & 5.0 for IBM AIX Power (64 bit) \\ 
 & & & & (November 26, 2003) \\
2 & tcmpibm & AIX-Power64 & 299056 & 5.0 for IBM AIX Power (64 bit) \\ 
 & & & & (November 26, 2003) \\
3 & tcmpibm & AIX-Power64 & 385182 & 5.0 for IBM AIX Power (64 bit) \\ 
 & & & & (November 26, 2003) \\
\hline
\end{tabular}
\end{center}
The results shown in this table are for the above three slaves named as
link1, link2 and link3 respectively, where all these slaves are opened
from the local computer named as `tcmpibm' (say). To get the information
about the total number of slaves those are opened, we use the command,

\vskip 0.2in
\begin{center}
{\fbox{\parbox{1.45in}{\centering{
Length[$\$$Slaves]
}}}}
\end{center}
For this case, the total number of slaves becomes $3$.

\section{How to Open Slaves in Remote Computers Available in Network ?}

To start a slave in remote computer, the command `ssh' is used which 
offers secure cryptographic authentication and encryption of the 
communication between the local and remote computer. Before starting 
a slave in a remote computer, it is necessary to check whether `ssh'
is properly configured or not, and this can be done by using the 
prescription,

\vskip 0.2in
\begin{center}
{\fbox{\parbox{1.75in}{\centering{
ssh remotehost math
}}}}
\end{center}
For example, if we want to connect a remote computer named as `tcmpxeon',
we should follow the command as,

\vskip 0.2in
\begin{center}
{\fbox{\parbox{1.75in}{\centering{
ssh tcmpxeon math
}}}}
\end{center}
Since `ssh' connection for a remote computer is password protected, it 
is needed to insert proper password, and if `ssh' is configured correctly,
the above operation shows the command `In[1]:='. Once `ssh' works 
correctly, a mathematica slave can be opened in a remote computer 
through this command,

\vskip 0.2in
\begin{center}
{\fbox{\parbox{4.75in}{\centering{
LaunchSlave[``remotehost", ``ssh -e none \`{}1\`{} math -mathlink"]
}}}}
\end{center}
For our illustrative purposes, below we describe how different slaves
with proper names can be started in different remote computers.

\vskip 0.2in
\begin{center}
{\fbox{\parbox{5.65in}{\centering{
link1=LaunchSlave[``tcmpxeon.saha.ac.in", ``ssh -e none \`{}1\`{} 
math -mathlink"] \\
link2=LaunchSlave[``tcmp441d.saha.ac.in", ``ssh -e none \`{}1\`{} 
math -mathlink"] \\
link3=LaunchSlave[``tcmpxeon.saha.ac.in", ``ssh -e none \`{}1\`{} 
math -mathlink"] \\
link4=LaunchSlave[``tcmp441d.saha.ac.in", ``ssh -e none \`{}1\`{} 
math -mathlink"] \\
}}}}
\end{center}
Here link1, link2, link3 and link4 are the four different slaves, where
the link1 and link3 are opened in a remote computer named as `tcmpxeon'
(say), while the other two slaves are started in another one remote
computer named as `tcmp441d' (say). Using this prescription, several
mathematica slaves can be started in different remote computers available 
in the network. The details of the above four slaves can be expressed
in the tabular form as,

\vskip 0.2in
\begin{center}
\begin{tabular}{|c|c|c|c|c|}
\hline
ID & host & OS & process & Version \\ \hline 
1 & tcmpxeon & Linux & 5137 & 5.0 for Linux (November 18, 2003) \\ 
2 & tcmp441d & Linux & 11323 & 5.0 for Linux (November 18, 2003) \\ 
3 & tcmpxeon & Linux & 5221 & 5.0 for Linux (November 18, 2003) \\ 
4 & tcmp441d & Linux & 11368 & 5.0 for Linux (November 18, 2003) \\ 
\hline
\end{tabular}
\end{center}
Thus we are now able to start mathematica slaves in local computer
as well as in remote computers available in the network, and with
this above background, we can describe the mechanisms for parallelizing
a mathematica program.

\section{Parallelizing of Mathematica Programs by using Remote Computers
Available in Network}

In order to understand the basic mechanisms of parallelizing a mathematica
program, let us begin with a very simple problem. We set the problem as
follows:

\vskip 0.15in
\noindent
{\bf\em\underline{Problem}}: {\em Construct a square matrix of any order
in a local computer and two other square matrices of the same order with
the previous one in two different remote computers. From the local computer,
read these two matrices those are constructed in the two remote computers.
Finally, take the product of these three matrices and calculate the
eigenvalues of the product matrix in the local computer.}

To solve this problem we proceed through these steps in a mathematica 
notebook. 

\vskip 0.15in
\noindent
{\em\underline{Step-1}} : For the sake of simplicity, let us first define
the names of the three different computers those are needed to solve this
problem. The local computer is named as `tcmpibm', while the names of the
other two remote computers are as `tcmpxeon' and `tcmp441d' respectively.
Opening a mathematica notebook in the local computer, let us first load
the package for parallelization, and to get the optional features, we load
another one package as mentioned earlier in Section 2. Then we start two 
mathematica slaves named as `link1' and `link2' in the two remote computers
`tcmpxeon' and `tcmp441d' respectively by using the proper commands as 
discussed in Section 3.

\vskip 0.15in
\noindent
{\em\underline{Step-2}} : Next we make ready three programs for the three
separate square matrices of same order in the local computer. Out of which 
one program will run in the local computer, while the rest two will run in
the two remote computers. These three programs are as follows.

\vskip 0.2in
\begin{center}
I. {\fbox{\parbox{5in}{\centering{
sample1$[$ns$_-]$$:=$Block$[\{$esi$=0, t=1.2, p=2.1$,vacuum1$=$$\{\}$$\}$,

Do[Do[a1=If[i$==$j,esi,0];

a2=If[$i<j$ $\&\&$ Abs$[i - j]$$==1, t, 0$];

a3=If[$i>j$ $\&\&$ Abs$[i - j]$$==1, p, 0$];

a4=a1+a2+a3;

a5 = AppendTo[vacuum1,a4], $\{j, 1, ns\}$], $\{i, 1, ns\}$];

a6 = Partition[a5, ns]$]$
}}}}
\end{center}

\vskip 0.2in
\begin{center}
II. {\fbox{\parbox{5in}{\centering{
sample2$[$ns$_-]$$:=$Block$[\{$esi$=0, q=2.6, r=1.8$,vacuum2$=$$\{\}$$\}$,

Do[Do[a1=If[i$==$j,esi,0];

a2=If[$i<j$ $\&\&$ Abs$[i - j]$$==1, q, 0$];

a3=If[$i>j$ $\&\&$ Abs$[i - j]$$==1, r, 0$];

a4=a1+a2+a3;

a5 = AppendTo[vacuum2,a4], $\{j, 1, ns\}$], $\{i, 1, ns\}$];

a6 = Partition[a5, ns]$]$
}}}}
\end{center}

\vskip 0.2in
\begin{center}
III. {\fbox{\parbox{5in}{\centering{
sample3$[$ns$_-]$$:=$Block$[\{$esi$=0, u=2, v=3$,vacuum3$=$$\{\}$$\}$,

Do[Do[a1=If[i$==$j,esi,0];

a2=If[$i<j$ $\&\&$ Abs$[i - j]$$==1, u, 0$];

a3=If[$i>j$ $\&\&$ Abs$[i - j]$$==1, v, 0$];

a4=a1+a2+a3;

a5 = AppendTo[vacuum3,a4], $\{j, 1, ns\}$], $\{i, 1, ns\}$];

a6 = Partition[a5, ns]$]$
}}}}
\end{center}
Since we are quite familiar about the way of writing mathematica 
programs~\cite{wolfram,san1}, we do not describe here the meaning of 
the different symbols used in the above three programs further. Thus 
by using these programs, we can construct three separate square 
matrices of order `ns'.

\vskip 0.15in
\noindent
{\em\underline{Step-3}} : We are quite at the end of our complete operation.
For the sake of simplicity, we assume that, the program-I is evaluated in
the local computer, while the program-II and program-III are evaluated in
the two remote computers respectively. All these three programs run 
simultaneously in three different computers. To understand the basic
mechanisms, let us follow the program.

\vskip 0.2in
\begin{center}
{\fbox{\parbox{5.75in}{\centering{
sample4$[$ns$_-]$$:=$Block$[\{\}$,

ExportEnvironment[``Global\`{}"];

mat1 = sample1[ns];

RemoteEvaluate[Export[``data1.dat", sample2[ns]], link1];

RemoteEvaluate[Export[``data2.dat", sample3[ns]], link2];

mat2 = RemoteEvaluate[ReadList[``data1.dat", Number, RecordLists$\rightarrow$ 
      True], link1];

mat3 = RemoteEvaluate[ReadList[``data2.dat", Number, RecordLists$\rightarrow$ 
      True], link2];

mat4 = mat1.mat2.mat3;

Chop[Eigenvalues[mat4]]$]$

}}}}
\end{center}
This is the final program. When it runs in the local computer, one matrix
called as `mat1' is evaluated in the local computer ($3$rd line of the 
program), and the other two matrices are determined in the remote computers
by using the operations given in the $4$th and $5$th lines of the program
respectively. The $2$nd line of the program gives the command for the 
transformations of all the symbols and definitions to the remote computers.
After the completion of the operations in remote computers, we call back
these two matrices in the local computer by using the command `ReadList', 
and store them in `mat2' and `mat3' respectively. Finally, we take the
product of these three matrices and calculate the eigenvalues of the 
product matrix in the local computer by using the rest operations of the
above program.

\vskip 0.12in
\noindent
The whole operations can be pictorially represented as,
\begin{figure}[ht]
{\centering \resizebox*{10.0cm}{6.5cm}{\includegraphics{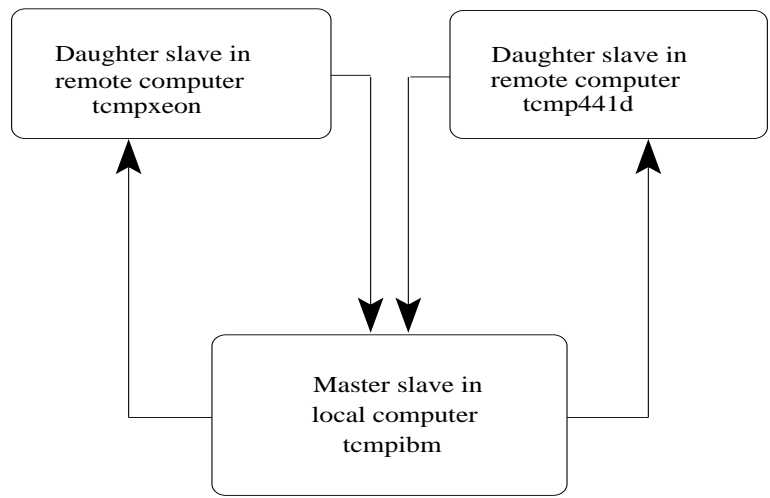}}\par}
\caption{Schematic representation of parallelization.}
\label{diagram}
\end{figure}

\vskip 0.12in
\noindent
At the end of all the operations, we close all the mathematica slaves by
using the following command.

\vskip 0.2in
\begin{center}
{\fbox{\parbox{1.25in}{\centering{
CloseSlaves[  ]
}}}}
\end{center}

\noindent
\addcontentsline{toc}{section}{\bf {Concluding Remarks}}
\begin{flushleft}
{\Large \bf {Concluding Remarks}}
\end{flushleft}
\vskip 0.1in
\noindent
In conclusion, we have explored the basic mechanisms for parallelizing 
a mathematica program by running its independent parts in remote computers
available in the network. By using this parallelization technique, one can 
enhance the efficiency of the numerical works, and it helps us to perform
all the mathematical operations within a very short period of time. In 
this present discussion, we have focused the parallelization technique for 
the Unix based operating system only. But all these operations also work 
very well in any other supported operating system like Windows, Macintosh, 
etc. Not only that, all these operations can also be done quite significantly
even if different versions of mathematica are installed in different remote 
computers those are used for parallel computation. 

\noindent
\addcontentsline{toc}{section}{\bf {Acknowledgment}}
\begin{flushleft}
{\Large \bf {Acknowledgment}}
\end{flushleft}
\vskip 0.1in
\noindent
I acknowledge with deep sense of gratitude the illuminating comments and
suggestions I have received from Prof. Sachindra Nath Karmakar during the
preparation of this article.

\addcontentsline{toc}{section}{\bf {References}}

\end{document}